\documentstyle[seceq,epsf,wrapft]{ptptex}

\pubinfo{Vol. 107, No. 6, June 2002}  

\notypesetlogo  

\title{
Instanton and Monopole in External Chromomagnetic Fields
}

\author{
Masahiro {\sc Fukushima},$^{1}$
Hideo {\sc Suganuma}$^{2}$
and 
Satoshi {\sc Chiba}$^{1}$
}

\inst{
$^{1}$Advanced Science Research Center, Japan Atomic Energy Research
Institute, Tokai 319-1195, Japan\\
$^{2}$Faculty of Science, Tokyo Institute of Technology, Tokyo 152-8551, Japan
}
\recdate{
December 7, 2001; Revised March 29, 2002}

\abst{
We study properties of instanton and monopole in an external 
chromomagnetic field. Generally, the 't~Hooft ansatz is 
no longer a solution of the Yang-Mills field equation 
in the presence of external fields. Therefore, we investigate a 
stabilized instanton solution with minimal total Yang-Mills 
action in a nontrivial topological sector. With this aim, we 
consider numerical minimization of the action with respect to
the global color orientation, the anisotropic scale 
transformation and the local gauge-like transformation starting 
from a simple superposed gauge field of the 't~Hooft ansatz 
and the external color field. Here, the external color field 
is, for simplicity, chosen to be a constant Abelian magnetic 
field along a certain direction. Then, the 4-dimensional rotational 
symmetry $O(4)$ of the instanton solution is reduced to two 
2-dimensional rotational symmetries $O(2)\times O(2)$ due 
to the effect of
a homogeneous external field. In the space $\mib{R}^{3}$ 
at fixed $t$, we find a quadrupole deformation of this 
instanton solution. In the presence of a magnetic field $\vec{H}$, 
a prolate 
deformation occurs along the direction of $\vec{H}$. 
Contrastingly, in the presence of an electric field $\vec{E}$
an oblate deformation occurs along the direction 
of $\vec{E}$. We further discuss 
the local correlation between the instanton and the 
monopole in the external field in the maximally Abelian gauge. 
The external field affects the appearance of the monopole 
trajectory around the instanton. In fact, a monopole and 
anti-monopole pair appears around the instanton center, and 
this monopole loop seems to partially screen the external field. 
}

\begin{document}
\maketitle

\section{Introduction}
\label{sec:sec1}
Quantum chromodynamics (QCD) includes various 
infrared phenomena, such as color confinement, 
dynamical chiral-symmetry breaking and large 
$\eta^{\prime}$ mass. It is necessary to 
clarify nonperturbative phenomena of the QCD vacuum 
composed of quarks, anti-quarks and gluon fields 
interacting in a highly complicated way. 
Topological properties
may provide a useful approach for descriptions of 
the nonperturbative nature of QCD, although the usual 
perturbative expansion is not applicable 
in this infrared region. The instanton solution 
\cite{Belavin:1975fg} is actually related to the 
$U_{A}(1)$ anomaly and large $\eta^{\prime}$ 
mass.\cite{'tHooft:1976up} Witten and Veneziano 
derived an approximate relation between the 
$\eta^{\prime}$ mass and topological susceptibility 
in the context of the $1/N_{c}$ 
expansion.\cite{Veneziano:1979ec,Witten:1979vv} Chiral symmetry 
breaking could also be interpreted as an instanton 
effect.\cite{Shuryak:1990cx,Schafer:1998wv,Diakonov:1995ea}

With recent progress in computational capabilities, the direct
investigation of instanton properties in the QCD 
vacuum using lattice QCD simulations has become practicable.\cite{Teper:1985rb,Ilgenfritz:1986dz,Polikarpov:1988yr,Campostrini:1990dh,Teper:1994un,Michael:1995br,deForcrand:1997sq,DeGrand:1997ss}
Elucidating instanton physics is important in order to 
understand the nonperturbative QCD vacuum.
A cooling procedure can be adopted to eliminate 
short-range quantum fluctuations and extract only 
topological excitations.\cite{Teper:1985rb}
Instantons seem to appear as deformed configurations 
even after cooling. The vacuum structure of 
non-Abelian gauge theories has been studied by Savvidy employing
the background field method.\cite{Savvidy:1977as} 
In the $SU(2)$ Yang-Mills theory, the 
effective potential up to one loop order for an Abelian gauge 
field leads to  a nonzero constant color magnetic field instead 
of the perturbative vacuum. Following Savvidy's work,
the Copenhagen group described the QCD vacuum including 
the effect of the inhomogeneous magnetic 
structure.\cite{Nielsen:1978rm,Ambjorn:1980ms,Ambjorn:1980xi}
Lattice QCD simulations yield an average instanton 
size $\bar{\rho} \simeq (0.33-0.4)~{\rm fm}$ and an
instanton number density $(N/V)\!\sim \langle G_{\mu\nu} 
G_{\mu\nu}/32\pi^{2}\rangle \simeq \!1~{\rm fm^{-4}}$,
corresponding to the gluon condensate.\cite{Teper:1985rb,Ilgenfritz:1986dz,Polikarpov:1988yr,Campostrini:1990dh,Teper:1994un,Michael:1995br,deForcrand:1997sq,DeGrand:1997ss,Negele:1998ev}
The QCD vacuum contains the gluon condensate and the 
inhomogeneous magnetic field. The 't~Hooft ansatz is 
not a solution of the Yang-Mills equation 
in the presence of such a background field, and therefore
the position, size, color orientation and 4-dimensional 
shape of each instanton are altered by the field.
In this paper, we study the instanton 
deformation mechanism by deriving a stabilized instanton 
solution that corresponds to a minimum Yang-Mills 
action in the external field. 

Recent studies have revealed the remarkable fact that instantons have 
a one-to-one correspondence with monopoles in the Abelian 
gauge, although these topological objects belong to different 
homotopy groups.\cite{Suganuma:1996qr,Miyamura:1995fc,Hart:1996wk,Fukushima:1997yn,Fukushima:1999gf,Brower:1997js}
This correspondence may provide a scenario of a color 
confinement mechanism that the instanton fields give rise to
the multi-production of monopole loops as a signal of 
monopole condensation.\cite{Fukushima:1997yn,Fukushima:1999gf}
For this reason, we further investigate the appearance of monopoles
around the instanton as a deformed configuration in external 
color magnetic fields.

\section{Instantons in an external field}
\label{sec:sec2}
The instanton is a classical nontrivial solution of the
Euclidean Yang-Mills field equations with a finite action,
which was discovered by Belavin, Polyakov, Schwartz 
and Tyupkin.\cite{Belavin:1975fg,Rajaraman:1982bk,Shuryak:1988bk}
The appearance of instantons corresponds to the homotopy 
group $\Pi_{3}(SU(N_{c}))=Z_{\infty}$.\cite{Rajaraman:1982bk}
In Minkowski space, instantons are
interpreted as tunneling events among degenerate vacua 
that are labeled by different winding numbers. The 
Euclidean Yang-Mills action is written
\begin{equation}
S  = \frac{1}{2g^{2}}\int d^{4}x {\rm tr}
\left( G_{\mu\nu} G_{\mu\nu} \right),  
\label{EYMA}
\end{equation}
where $G_{\mu\nu}\!\equiv [D_{\mu},D_{\nu}]\!=
\partial_{\mu}A_{\nu}\!- \partial_{\nu}A_{\mu}\!- [A_{\mu},A_{\nu}] $ 
is the field strength.\cite{Rajaraman:1982bk}
Here, to simplify the notation, the field strength and the 
gauge field are defined as $G_{\mu\nu} \equiv 
iG_{\mu\nu}^{a}\tau^{a}/2$ and $A_{\mu} \equiv iA_{\mu}^{a}
\tau^{a}/2$. By defining the dual field strength
as $\tilde{G}_{\mu\nu} \equiv \frac{1}{2} \varepsilon_{\mu\nu\alpha\beta} 
G_{\alpha\beta}$, we can rewrite the action as
\begin{equation}
S=\frac{1}{8g^{2}}\int d^{4}x \{ G^{a}_{\mu\nu} \pm 
\tilde{G}^{a}_{\mu\nu} \}^{2} \mp \frac{8\pi^{2}}{g^{2}} {\cal Q}.   
\end{equation}
From this formula, the action provides a minimal value 
characterized by the topological charge 
${\cal Q} \equiv (1/32\pi^{2}) \int d^{4}x G^{a}_{\mu\nu} 
\tilde{G}^{a}_{\mu\nu}$ when
the field strength satisfies the (anti)self-dual condition 
$G^{a}_{\mu\nu} = \pm \tilde{G}^{a}_{\mu\nu}$. This self-dual
relation satisfies the Yang-Mills field equation 
automatically, as
$[D_{\mu},G_{\mu\nu}] = \pm [D_{\mu},\tilde{G}_{\mu\nu}] 
= 0$, due to the Jacobi identity. 

The instanton solution is well known as the 't~Hooft 
ansatz.\cite{Rajaraman:1982bk,Shuryak:1988bk}
The single instanton solution with ${\cal Q}=1$ is written 
\begin{equation}
A^{I (s)}_{\mu}(x) = \frac{2iO^{ab}
\bar{\eta}^{b\mu\nu}(x-z)_{\nu}\rho^{2}}
{(x-z)^{2}\{ (x-z)^{2}+\rho^{2}\} }  \frac{\tau^{a}}{2}
\label{ETA}
\end{equation}
in the singular gauge.
The $SU(2)$ instanton has 8 collective 
coordinates corresponding to the size $\rho$, the 
4-dim central position $z_{\mu}$, and the global 
orientation $O^{ab}$ in color space. Here, 
$\eta_{b\mu\nu}$ denotes the 't~Hooft symbol, 
\cite{Rajaraman:1982bk} defined as 
\begin{eqnarray}
\bar{\eta}^{b\mu\nu} = - \bar{\eta}^{b\nu\mu} \equiv \left\{
\begin{array}{ll}
\varepsilon^{b\mu\nu}~,   &\quad (\mu,\nu = 1,2,3) \\
    -\delta^{b\mu}~,      &\quad (\nu     = 4)
\end{array}\right.
\end{eqnarray}
with $\eta_{b\mu\nu}\equiv (-1)^{\delta^{\mu 4}+
\delta^{\nu 4}}\bar{\eta}_{b\mu\nu}$.
The anti-instanton with ${\cal Q}=-1$ can 
be obtained by replacing $\bar{\eta}_{a\mu\nu}$ with 
$\eta_{a\mu\nu}$ in Eq.~(\ref{ETA}). The single instanton 
action density has a point-like peak at its center with 
the 4-dimensional rotational symmetry $O(4)$ in the 
space-time $\mib{R}^{4}$. 

The QCD vacuum contains, for instance, instanton and 
anti-instanton fields, the gluon condensate and inhomogeneous 
magnetic fields, as in the Copenhagen vacuum, and so on. 
Therefore, we would like to study the effects of background 
fields on classical configurations like instantons.
For simplicity, it is convenient to consider a translationally 
invariant external field. In the $SU(2)$ case, 
there are two categories of such external fields. By employing 
a suitable gauge, one is reduced to a QED-like case, and the 
other is reduced to a constant $A^{1}_{\mu}$ and $A^{2}_{\nu}$ case, 
which lead to a constant field strength $F^{3}_{\mu\nu}
\propto [A^{1}_{\mu},A^{2}_{\nu}]$. In this work, we consider 
the QED-like case as an idealized external field. We use an
external gauge field given by
\begin{equation}
A_{\mu}^{3 ex}=-\frac{1}{2}F_{\mu\nu}x_{\nu},~~~
A_{\mu}^{\pm ex}\equiv\frac{1}{\sqrt{2}}(A_{\mu}^{1}\pm iA_{\mu}^{2})=0,
\end{equation}
which leads to a constant Abelian field strength $G_{\mu\nu} 
= iF_{\mu\nu}\frac{\tau^{3}}{2}$. Here, $F_{\mu\nu}\in \mib{R}$ 
is a constant anti-symmetric tensor. 

In the presence of an external field, the 't~Hooft 
ansatz is no longer a solution of the Yang-Mills field 
equation. Therefore, we would like to find the instanton 
solution that minimizes the total action in external 
fields. First of all, we construct the total gauge field 
as a superposition of the instanton field and the 
external field: 
\begin{equation}
A^{tot}_{\mu}(x)=A^{ex}_{\mu}(x)+A^{I}_{\mu}(x). 
\end{equation}
The total Yang-Mills action given by this simple 
superposed configuration would be shifted from the absolute 
minimum due to the overlapping between $A^{ex}_{\mu}$ 
and $A^{I}_{\mu}$. Therefore, we derive a stabilized 
solution with nontrivial topological sector that
minimizes the total Yang-Mills action by employing a
variational analysis with respect to the instanton 
configuration $A^{I}_{\mu}$. 

\section{Minimization of total instanton action under
an external field}
\label{sec:sec3}
As the degrees of freedom of the variational space, 
we consider (i) the 
global color orientation $O^{ab}$, (ii) the 
anisotropic scale transformation $\lambda_{\mu}$, 
and (iii) the local gauge-like transformation 
$\Omega^{a}(x)$ of the instanton configuration. 
Certainly, the action of an isolated single instanton is 
independent of $O$ and $\Omega(x)$ due to the gauge 
invariance. However, the total action depends 
directly on these variables in the presence of 
external fields. We should start from the global 
variables and then perform the minimization locally 
in order to find the position of absolute minimum
in this variational space. In the actual calculation, 
we determine $O^{ab}$, $\lambda_{\mu}$ and $\Omega(x)$ 
in turn by minimizing the total Yang-Mills action with 
respect to each.

\subsection{Global color orientation}
First, we minimize the total action with respect to 
the global color orientation $O^{ab}$. 
A single instanton has 
collective coordinates concerning the global rotation 
in the $SU(2)$ color space given by
\begin{eqnarray}
O \!\equiv\!
\left[\!\!
\begin{array}{lllll}
-\!\sin\psi\sin\phi \!+\!\cos\psi\cos\theta\cos\phi \!\!\!\!&
~~\sin\psi\cos\phi \!+\!\cos\psi\cos\theta\sin\phi \!\!\!&
-\!\cos\psi\sin\theta \\
-\!\cos\psi\sin\phi\!-\!\sin\psi\cos\theta\cos\phi \!\!\!\!&
~~\cos\psi\cos\phi\!-\!\sin\psi\cos\theta\sin\phi \!\!\!&
~~\sin\psi\sin\theta\\
~~\sin\theta\cos\phi\!\!\!\!&
~~\sin\theta\sin\phi\!\!\!&
~~\cos\theta
\end{array}\!\! \right], \nonumber\\
\label{EGCO}
\end{eqnarray}
which is characterized by the three Euler angles $\theta \in 
[0,\pi]$ and $\phi, \psi  \in [0,2\pi)$. 
In the absence of external fields, the global $O(4)$ 
symmetry is manifest, and these variables are irrelevant 
to the action. However, if there is an external field 
breaking the 4-dimensional space-time symmetry, these 
Euler angles would be determined from the condition of 
a minimal total action.

\subsection{Anisotropic scale transformation}
As the next step, we carry out an anisotropic scale 
transformation of the instanton configuration as
\begin{equation}
x_{\mu} \rightarrow \lambda_{\mu} x_{\mu}.
\label{EAS}
\end{equation}
Here, $\lambda_{\mu}$ are the global rescale 
parameters acting on the coordinates $x_{\mu}$ 
along the direction $\mu$.
This anisotropic scale transformation directly 
affects the instanton profile, which is not done
by a transformation of the gauge degrees of 
freedom of the instanton configuration. 
Here, we impose the condition $\lambda_{1} \lambda_{2} 
\lambda_{3} \lambda_{4} = 1$ and fix the 4-dimensional 
volume of the instanton profile in terms of the 
characteristic scale $\bar{\rho}$. At this point, 
we should give a comment on the size of an instanton. 
Generally, the overlapping between an instanton and the 
background field increases the total action density. A 
larger size of the instantons leads to a much 
larger increase of the action, and the stability of 
a finite size instanton is not certain in a uniform 
external field. However, 
in the QCD vacuum,  instantons and anti-instantons of a
characteristic size and density have been observed using 
lattice QCD simulations.\cite{Teper:1985rb,Ilgenfritz:1986dz,Polikarpov:1988yr,Campostrini:1990dh,Teper:1994un,Michael:1995br,deForcrand:1997sq,DeGrand:1997ss,Negele:1998ev}
It has been found also in the simplified 
instanton gas model that the instanton size distribution 
has a peak at a characteristic instanton size.\cite{Munster:2000uu} 
With these results in mind, we here impose the above condition 
by hand and consider 
only the deformation effect on an instanton in such a 
characteristic size. 
The breaking of the $O(4)$ symmetry due to external fields 
leads to a nontrivial condition of $\lambda_{\mu}$, which 
minimize the total action.

\subsection{Local gauge-like transformation}
Until this point, we have considered only the global 
parameters of the instanton 
configuration. The self-duality of an instanton solution is 
generally broken in the presence of an external field, and 
the true minimum 
of the total action cannot be realized through a global
transformation alone for such a self-dual initial configuration.
For this reason, we also consider the local minimization 
of the total action considering gauge-like degrees of freedom 
of the instanton configuration. We formulate the gauge-like 
transformation as follows. In the construction of the total 
gauge fields, there is an ambiguity of the ``gauge-like 
choice'' of $A^{I}_{\mu}(x)$ as
\begin{eqnarray}
A^{I^{\prime}}_{\mu}(x) = \Omega(x)( A^{I}_{\mu}(x)
+\partial_{\mu} ) \Omega^{\dagger}(x), \label{ELGLT1}
\end{eqnarray}
in the presence of an external $A^{ex}_{\mu}(x)$,
whose gauge degrees of freedom are fixed as
\begin{eqnarray}
A^{ex^{\prime}}_{\mu}(x) = A^{ex}_{\mu}(x).
\label{ELGLT2}
\end{eqnarray}
For instance, the instanton gauge field $A^{I sing}_{\mu}$ 
in the singular gauge can be expressed equivalently as 
\begin{equation}
A^{I reg}_{\mu}= \Omega^{sing}( A^{I sing}_{\mu}
+\partial_{\mu} ) \Omega^{\dagger sing},
\label{EGT}
\end{equation}
in the regular gauge with the singular gauge function 
$\Omega^{sing}= (x_{4}+x_{i}\tau_{i})/|x|$. 
Certainly, the gauge invariant quantities, like the 
instanton action $S^{I}$ and topological charge 
${\cal Q}^{I}$, are independent of this gauge choice
in the absence of external fields. However, external 
fields cause the total action $S^{tot}$ to depend on 
the ``gauge-like choice'' of the instanton part 
$A^{I}_{\mu}$, like Eq.~(\ref{EGT}), although 
$S^{tot}$ is, of course, independent of the ``total'' 
gauge transformation of $A^{tot}_{\mu}=A^{ex}_{\mu}+A^{I}_{\mu}$ 
given by 
\begin{equation}
{A^{tot}_{\mu}}^{\prime}(x)= \Omega(x)( A^{tot}_{\mu}(x)
+\partial_{\mu} ) \Omega^{\dagger}(x).
\label{SGT}
\end{equation}
We call this partial gauge transformation of Eqs.~(\ref{ELGLT1}) 
and (\ref{ELGLT2}) a ``gauge-like'' transformation. Such 
a ``gauge-like'' function should be determined iteratively 
by the local minimization of the total action $S^{tot}$. 

\section{Lattice formulation}
\label{sec:sec4}
The lattice technique is applicable to the local 
minimization of the total Yang-Mills action for the 
gauge-like function $\Omega(x)$. Using the lattice, 
we can additionally employ the maximally Abelian (MA) 
gauge fixing and extract monopole currents as will 
be considered in \S\ref{sec:sec6}. The link variables 
both for the instanton and the external field can be 
defined as 
\begin{eqnarray}
U_{\mu}^{ I}(s) &\equiv& \exp [ ia A^{I }_{\mu}(s)], \\
U_{\mu}^{ex}(s) &\equiv& \exp [ ia A^{ex}_{\mu}(s)], 
\end{eqnarray}
respectively. We use the standard Wilson action as the
lattice Yang-Mills action, which can be rewritten as 
\begin{equation}
S^{tot}_{YM}=\frac{1}{8\pi^{2}}\sum_{s,|\mu|>|\nu|,(ij) 
} \!{\rm tr}\!\Bigm[ 1-P^{(ij)}_{\mu\nu}(s)\Bigm] +~{\rm h.c.},
\end{equation}
with the plaquette variable 
\begin{equation}
P^{(ij)}_{\mu\nu} \equiv 
U^{i}_{\mu}(s) U^{j}_{\nu}(s+\hat{\mu}) 
U^{i\dagger}_{\mu}(s+\hat{\nu}) U^{j\dagger}_{\nu}(s).
\end{equation}
Here, the summation over $\mu$ and $\nu$ is taken as
$\mu,\nu \in \{ \pm 1,\pm 2,\pm 3,\pm 4 \}$.
With regard to the preservation of 4-dim geometrical symmetry,
this clover-type action seems preferable. 
The labels $i,j \in \{ R,L\}$ relate to the ordering 
ambiguity in the definition of the total link variables 
due to the lattice discretization. In fact, we have 
the two choices 
\begin{eqnarray}
U^{R}_{\mu}(s)&\equiv&
U^{ex}_{\mu}(s)U^{I}_{\mu}(s),\\
U^{L}_{\mu}(s)&\equiv& 
U^{I}_{\mu}(s)U^{ex}_{\mu}(s), 
\end{eqnarray}
which reproduce 
$A^{tot}_{\mu}(x)=A^{ex}_{\mu}(x)+A^{I}_{\mu}(x)$
in the continuum limit. We actually combine
$U^{R}_{\mu}$ and $U^{L}_{\mu}$ in such a manner to 
maintain the geometrical symmetry and the ordering 
symmetry for the product of $U^{I}_{\mu}$ and 
$U^{ex}_{\mu}$, as shown in the Appendix. 

Under the anisotropic scale transformation following
Eq.~(\ref{EAS}), the instanton link variables become
\begin{equation}
U^{I}_{\mu}(\lambda_{\alpha} x_{\alpha}) 
\equiv \exp\bigm( i \lambda_{\mu} a
A^{I}_{\mu}(\lambda_{\alpha} x_{\alpha})\bigm).  
\label{EAST}
\end{equation}
Here, it is noted that the integral elements $a$ 
are replaced with $\lambda_{\mu} a$. 

The local gauge-like transformation 
of Eqs.~(\ref{ELGLT1}) and (\ref{ELGLT2}) at a site $s_{0}$ 
is defined on the lattice as
\begin{eqnarray}
U^{I^{\prime}}_{\mu}(s_{0})&=&\Omega(s_{0})
U^{I}_{\mu}(s_{0})\Omega^{\dagger}(s_{0}+\hat{\mu}),\\
U^{ex^{\prime}}_{\mu}(s_{0})&=&U^{ex}_{\mu}(s_{0}),
\end{eqnarray}
with $\Omega(s_{0})=\Omega_{0}(s_{0}){\bf 1}+i\Omega_{i}(s_{0}) 
\tau_{i}$, while $U^{I}_{\mu}(s)$ and $U^{ex}_{\mu}(s)$ are fixed 
at the sites $s \ne s_{0}$. Under the local gauge-like transformation 
$\Omega(s_{0})$, the Wilson action density at $s_{0}$ changes as
$S_{YM}(s_{0}) \rightarrow S^{\prime}_{YM}(s_{0})$, with
$S^{\prime}_{YM}(s_{0}) = S_{YM}(s_{0})-\delta S_{YM}(s_{0};\Omega(s_{0})).$
Apart from an irrelevant constant, the difference of the total 
action density, $\delta S_{YM}$, can be written in a 
bilinear form of $\Omega$,
\begin{eqnarray}
\delta S_{YM}(s_{0};\Omega(s_{0}))
= \sum_{\mu>\nu} \sum^{12}_{\alpha=1} 
{\rm tr} \Bigm[ \hat{L}^{\alpha}_{\mu\nu}
\Omega\bar{L}^{\alpha}_{\mu\nu}
\Omega^{\dagger}(s_{0})\Bigm],
\label{DIFFAC}
\end{eqnarray}
where $\hat{L}^{\alpha}_{\mu\nu}$ and $\bar{L}^{\alpha}_{\mu\nu}$ 
are given by $U^{I}$ and $U^{ex}$ as shown in the Appendix. 
At the boundary of the $N^{4}$ 
lattice, we employ a trivial boundary as $\Omega^{0}(s) 
= 1$ and $\Omega^{i}(s)=0$. We can determine the appropriate
function $\Omega(s)$ by maximizing $\delta S$ at each 
site, and we obtain the instanton solution with minimal
total action iteratively.

\section{Numerical results}
\label{sec:sec5}
In the actual calculation, we considered a typical size 
instanton with $\rho=0.4~{\rm fm}$.~\cite{Shuryak:1988bk}
Although a single instanton possesses scale invariance in 
the continuum space, a small-size instanton with $\rho<a$ 
cannot be described on a lattice because of the finite 
lattice spacing $a$. In order to guarantee lattice continuity 
for instanton configurations, we used a lattice spacing 
$a=0.05~{\rm fm}$ that is quite fine compared to the 
instanton size $\rho=0.4~{\rm fm}$
on a $N^{4} =32^{4}$ lattice. From the instanton liquid 
model,\cite{Shuryak:1988bk} 
we have another scale parameter, the 
gluon condensate, $\frac{1}{32\pi^{2}} \langle G^{a}_{\mu\nu}
G^{a}_{\mu\nu} \rangle \simeq (200~{\rm MeV})^{4}$. As a 
typical magnitude of external fields, $F_{\mu\nu}$ is set to 
be on the order of the gluon condensate. Without loss of 
generality, we take $F_{12} > 0$ and set the other components 
equal to zero in a suitable Lorentz frame. This case 
corresponds to that of a constant magnetic system. 

We adopt the 't~Hooft ansatz in the singular gauge
as the initial instanton configuration superposed on the 
external field. Although we can also use the regular gauge 
for the instanton, the regular gauge function spreading 
over a wide region leads to a larger increase in the action due 
to the overlapping with the external field than in the 
case of the singular gauge. This is the practical reason why 
we use the singular 
gauge initially. Certainly, we would obtain the same result
starting from different gauge choices.

\begin{wrapfigure}{r}{\halftext}
\epsfxsize = 6.6cm
\centerline{\epsfbox{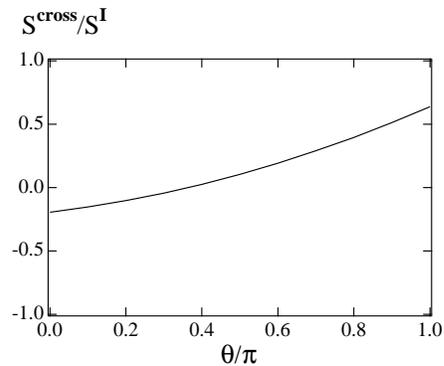}}
\vspace*{-0.6cm}
\caption{The dependence of the total action on the 
global color-orientation angle $\theta$ in a
homogeneous external field with $F_{12} > 0$.}
\label{FCO}
\end{wrapfigure}

Figure~\ref{FCO} shows the dependence of the total action on
the global color orientation of Eq.~(\ref{EGCO}). Here, 
we plot contribution from the cross term between 
the instanton and external field to the total action,
 defined as
\begin{equation}
S^{cross}/S^{I} \equiv ( S^{tot}-S^{I}-S^{ex}  )/S^{I},
\end{equation}
which is normalized by the isolated instanton action 
$S^{I}$. Here, $S^{tot}$ and $S^{ex}$ denote the total 
Yang-Mills action and the external action, respectively. 
This variable $S^{cross}$ differs from $S^{tot}$ only 
by a constant, since $S^{I}$ and $S^{ex}$ are constants.

The total action is independent both of the angles $\psi$
and $\phi$ in Eq.~(\ref{EGCO}), because of the $O(2)$ symmetry 
along the direction of the external field. In sharp contrast,
the total action depends rather strongly on $\theta$, as shown 
in Fig.~\ref{FCO}. We find the value $\theta =0$ that 
minimizes the total action for the global color 
orientation. In the absence of external fields, these parameters 
of the single instanton are collective coordinates, 
and cannot be determined uniquely. 

\begin{figure}[h]
\parbox{\halftext}{
\epsfxsize = 6.5cm
\centerline{
\epsfbox{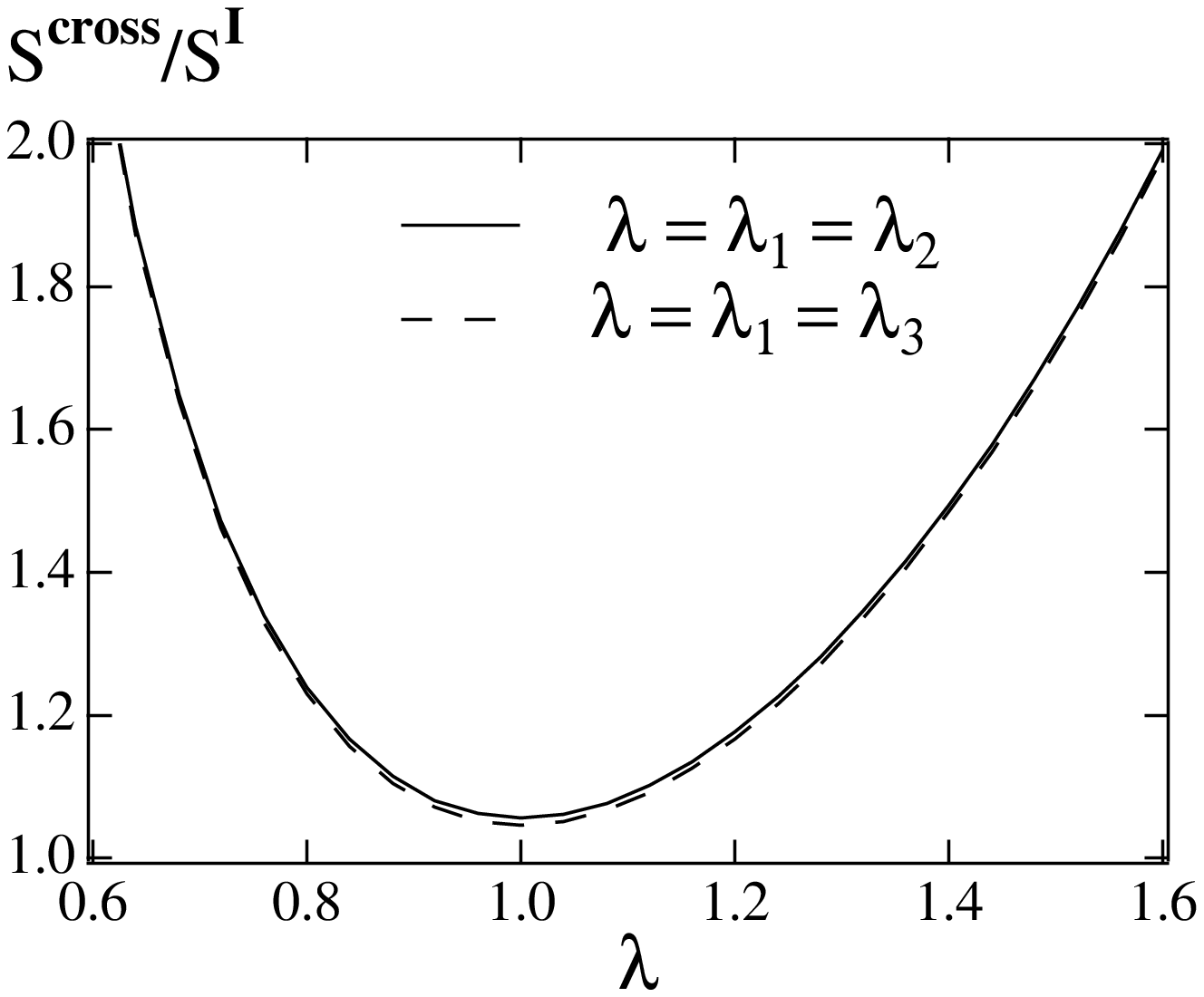}}
\vspace*{-0.6cm}
\center{(a)}}
\hfill
\parbox{\halftext}{
\epsfxsize = 6.5cm
\centerline{
\epsfbox{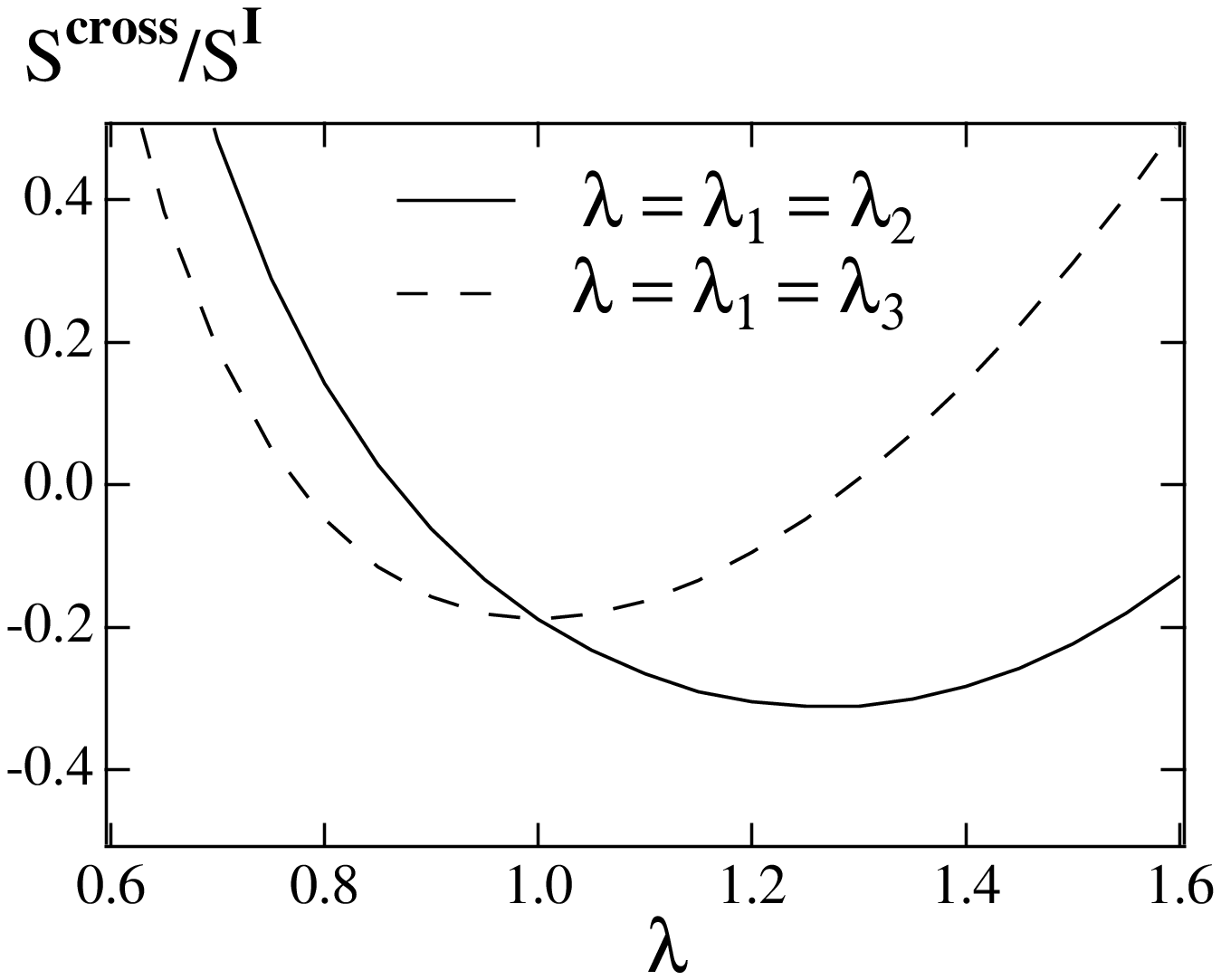}}
\vspace*{-0.6cm}
\center{(b)}}
\caption{The dependence of the total action 
on the anisotropic scale transformation with 
$\lambda_{\mu}$. The solid curve represents the result 
with fixed $\lambda_{1}=\lambda_{2}=\lambda$ 
and $\lambda_{3}=\lambda_{4}=1/\lambda$, and 
the dashed curve represents that 
$\lambda_{1}=\lambda_{3}=\lambda$ 
and $\lambda_{2}=\lambda_{4}=1/\lambda$. 
(a) The isolated instanton case without 
an external field. (b) The interacting instanton 
case with an external field $F_{12}>0$.}
\label{FAS1}
\end{figure}

After determining the global angles, we 
carried out the anisotropic transformation of the instanton 
profile given by Eq.~(\ref{EAST}). It is instructive to consider 
the case without an external field, in which the instanton is 
a classical solution of the Yang-Mills field equation. An 
arbitrary anisotropic scale transformation leads to 
an increase of the action, where the minimal action condition 
is trivial at $\lambda_{1}=\lambda_{2}=\lambda_{3}=
\lambda_{4}=1$, as shown in Fig.~\ref{FAS1}(a). Thus, 
the single instanton is found to be stable with
4-dimensional spherical symmetry.

\begin{wrapfigure}{l}{\halftext}
\epsfxsize = 6.6cm
\centerline{\epsfbox{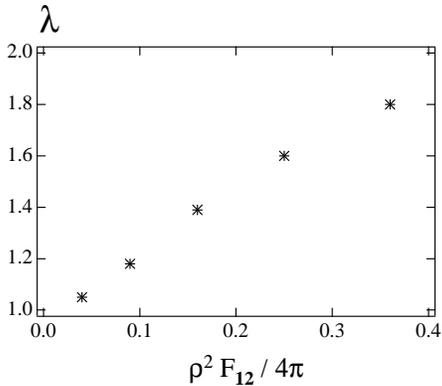}}
\vspace*{-0.3cm}
\caption{The dependence of the magnitude of the 
instanton deformation on the size $\rho$ 
and the strength of the external field $F_{12}$. 
We plot the parameter $\lambda$ that provides
the minimal Yang-Mills action as a function 
of the dimensionless variable $\rho^{2} 
F_{12}/4\pi$.}
\label{FAS2}
\end{wrapfigure}

The external field leads to a deformation 
of the instanton profile. Here, it is noted that 
the 4-dimensional rotational symmetry $O(4)$ is reduced 
to two 2-dimensional rotational symmetries $O(2)\times 
O(2)$, due to the homogeneous external field $F_{12}> 0$. 
The scale transformation with fixed $\lambda_{1}=\lambda_{2}$ 
and $\lambda_{3}=\lambda_{4}$ provides a nontrivial minimum 
point, as shown in Fig.~\ref{FAS1}(b), although the total 
action tends to increase under the deformation with fixed
$\lambda_{1}=\lambda_{3}$ and $\lambda_{2}=\lambda_{4}$.
In this external field, the instanton is stable for a 
quadrupole-deformed configuration in $\mib{R}^{3}$. From 
this result, we  determine the parameters $\lambda_{\mu}$ as
$\lambda_{1}=\lambda_{2}=\lambda=1.3$ and $\lambda_{3}=
\lambda_{4}=1/\lambda$ to realize the minimal total action 
with respect to the anisotropic transformation.
Figure~\ref{FAS2} shows that both the instanton size 
and the strength of the external field affect the
deformation of the instanton profile. 
The parameter $(\lambda-1)$ is approximately proportional 
to the dimensionless parameter $\rho^{2}F_{12}$. 
It is found that the instanton would be deformed more 
strongly by the background fields as the size increases. 

\begin{figure}[t]
\parbox{\halftext}{
\epsfxsize = 6.6cm
\centerline{\epsfbox{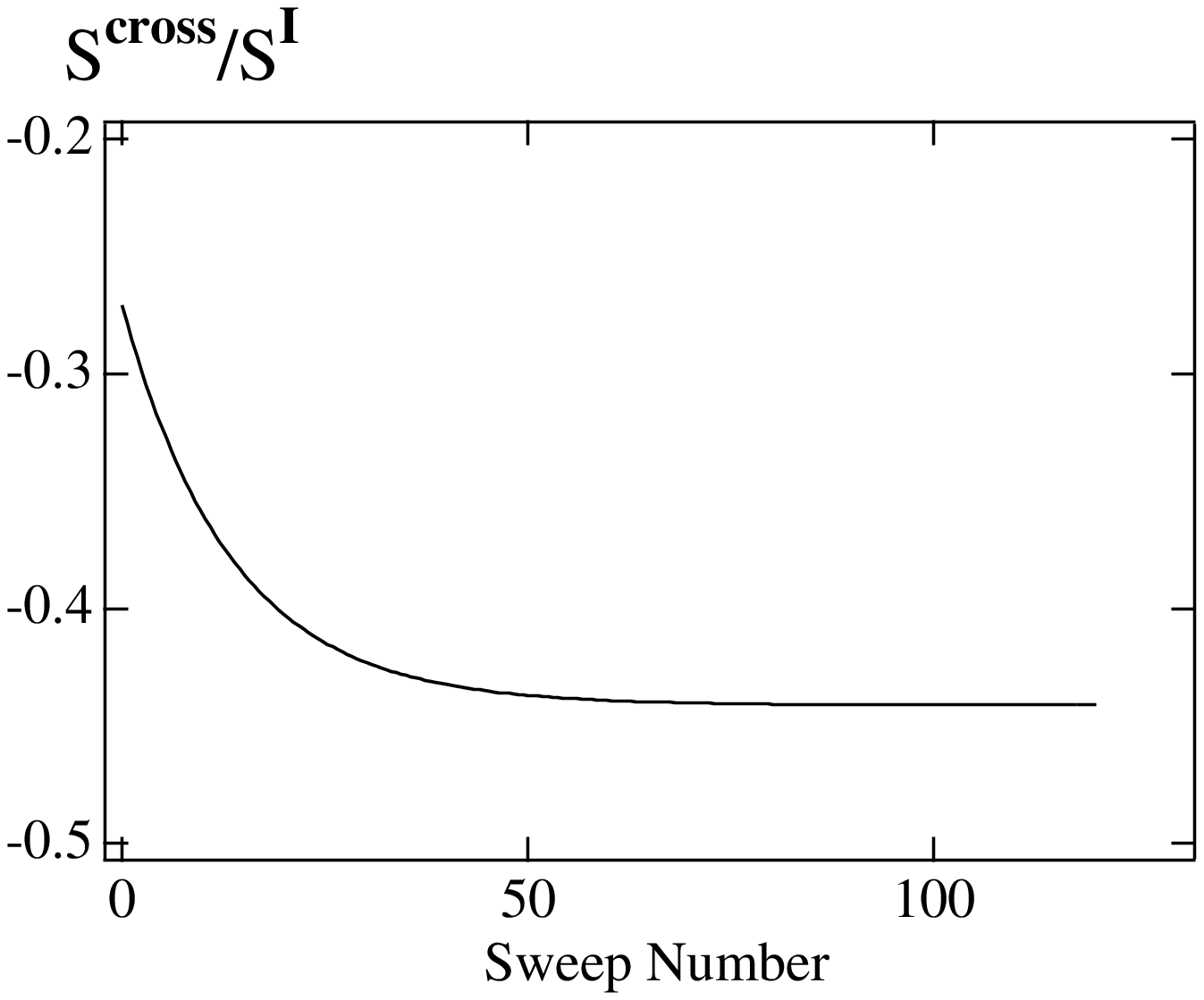}}
\vspace*{-0.6cm}
\caption{The cooling curve of the total action
for the local gauge-like transformation of
$A^{I}_{\mu}(\theta_{min},\lambda_{min})$. 
The initial configuration is the 't Hooft ansatz
in the singular gauge with $\theta_{min}=0$ and 
$\lambda_{min}=1.3$. }
\label{FGL}
}
\hfill
\parbox{\halftext}{
\epsfxsize = 5cm
\centerline{\epsfbox{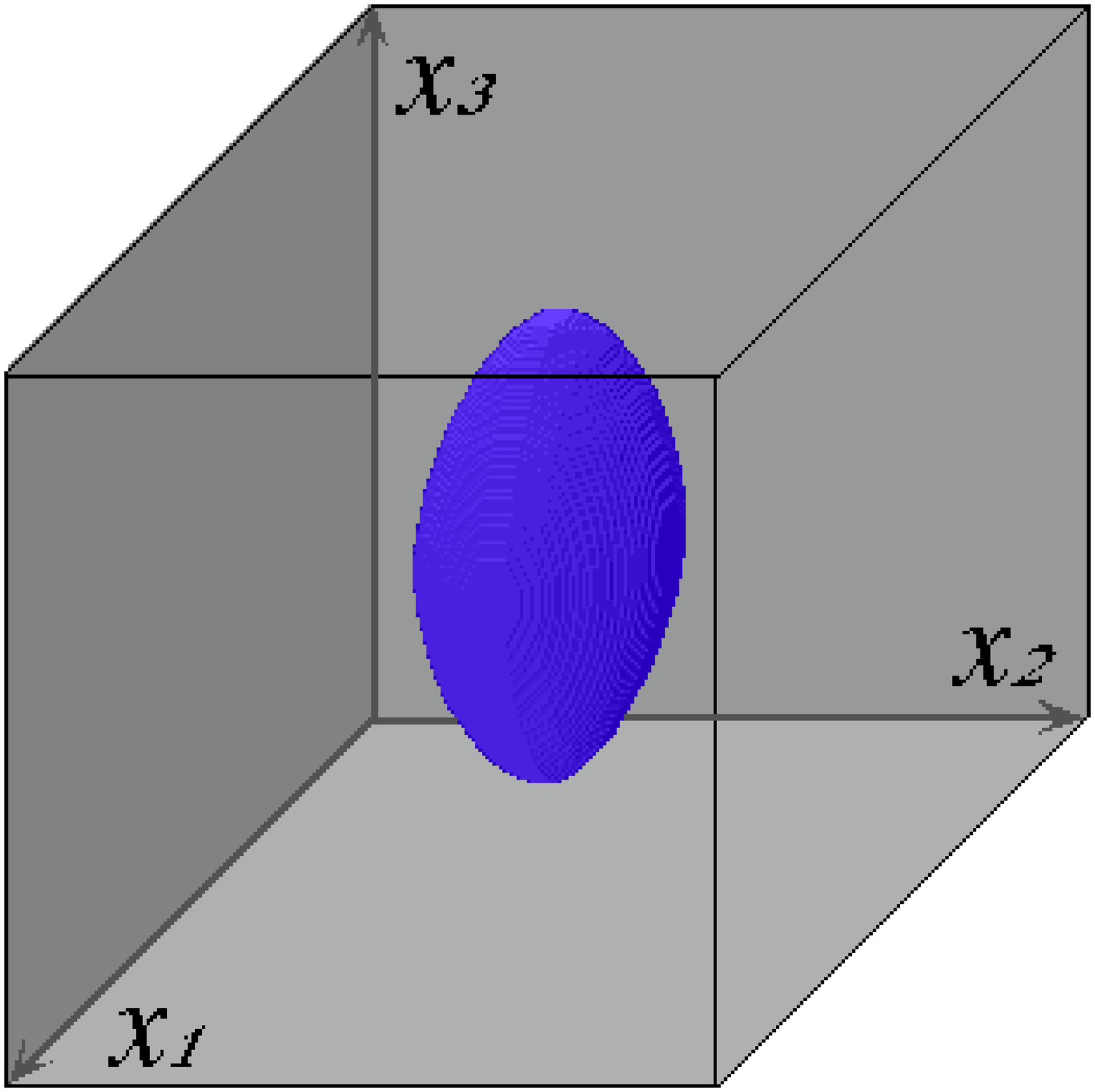}}
\caption{The total action density of the 
instanton solution in an external field 
in a cross section of $\mib{R}^{3}$
in the $t$ plane, including the instanton 
center. The presently considered external 
field has a nonzero magnetic component 
$H_{3}$ along the third direction, $x_{3}$.}
\label{FFR}}
\vspace{-0.3cm}
\end{figure}

To this point, we have determined the global color orientation 
and the anisotropic scale transformation as global 
conditions. In order to obtain a stable instanton solution 
in the external field, we further consider the gauge-like 
transformation of Eqs.~(\ref{ELGLT1}) and (\ref{ELGLT2}) 
as a local minimization of the total Yang-Mills action. We 
plot the cooling curve for the local gauge-like 
transformation in Fig.~\ref{FGL}. The initial 
configuration in Fig.~\ref{FGL} was determined previously to 
be the 't~Hooft ansatz of Eq.~(\ref{ETA}) in the singular gauge 
with the global parameter values $\theta_{min} =0$ and 
$\lambda_{1}=\lambda_{2}=\lambda_{min}=1.3$. With this 
iteration, the total action further decreases and converges 
after about 100 sweeps. Consequently, we obtain the stable
instanton solution with minimal total Yang-Mills action, 
which cannot be obtained by considering only global parameters 
like $O^{ab}$ and $\lambda$. Finally, we note that the 
instanton seems to be a quadrupole-deformed configuration 
in $\mib{R}^{3}$, as shown in Fig.~\ref{FFR}. 

To summarize this section, we considered the instanton deformation 
mechanism in a homogeneous external color field. In an
external field with $F_{12}> 0$, the total action depends 
strongly on the angle $\theta$ of the global color orientation 
and the parameters $\lambda_{\mu}$ of the anisotropic scale 
transformation. We obtained the minimal total Yang-Mills action
at $\theta=0$ and $\lambda_{1}=\lambda_{2}=1/\lambda_{3}=1/
\lambda_{4}>\lambda = 1.3$. The magnitude of the instanton 
deformation depends both on the instanton size and the 
strength of the external field. A large instanton seems to be affected 
and deformed strongly by background fields.  Finally, we obtained
a stable instanton solution where the $O(4)$ symmetry is reduced 
to two 2-dimensional rotational symmetries $O(2)\times 
O(2)$ in the $xy$ and $zt$ planes, due to the homogeneous external 
field with $F_{12} = H_{3}> 0$. In $\mib{R}^{3}$ at a fixed $t$, 
we find a quadrupole deformation of this instanton solution. In 
presence of a magnetic field $\vec{H}$, a prolate deformation occurs 
along the direction of $\vec{H}$. In an electric field $\vec{E}$, 
which corresponds to e.g. $F_{14}\ne 0$, an oblate deformation 
occurs along the direction of $\vec{E}$. 
Due to the external field, the instanton solution loses its
4-dimensional rotational symmetry.

\section{Monopole around an instanton in an external field}
\label{sec:sec6}
In 1981, 't~Hooft proposed the Abelian gauge,\cite{'tHooft:1981ht}
for which the $SU(N_{c})$ non-Abelian gauge theory is 
reduced to the $U(1)^{N_{c}-1}$ Abelian gauge theory 
with color-magnetic monopoles. The appearance of magnetic 
monopoles corresponds to the homotopy group $\Pi_{2}(SU(N_{c})
/U(1)^{N_{c}-1})=Z_{\infty}^{N_{c}-1}$, 
which is different from that of instantons. In the Abelian 
gauge, the color-magnetic monopole appears as a relevant 
degree of freedom for the description of color confinement,
which was proposed by Nambu, 't Hooft and Mandelstam in the 
mid-1970s.\cite{Nambu:1974pb,Nambu:1974zg,'tHooft:1975pu,Mandelstam:1974pi}
This mechanism can be interpreted as the dual Meissner effect 
due to monopole condensation, which is a dual version of 
Cooper pair condensation in ordinary superconductivity. 
This analogy is based on the duality between the magnetic 
and electric parts. In lattice QCD simulations, it has been 
observed that large monopole clustering covers the entire physical 
vacuum in the confinement phase in the MA gauge. This has been
cited as evidence that monopole condensation is
responsible for confinement.\cite{Kronfeld:1987vd,Kronfeld:1987ri}
Many studies indicate that the monopole is a relevant degree of 
freedom for color confinement and chiral symmetry breaking.\cite{Brandstater:1991sn,Hioki:1991ai,Kitahara:1995vt,Miyamura:1995xn,Miyamura:1994xk,Woloshyn:1995rv,Suganuma:1995ps,Sasaki:1995sa,Sasaki:1996ss}

Recent studies show that monopoles are directly related 
to instantons, although these topological objects
belong to different homotopy groups.\cite{Suganuma:1996qr,Miyamura:1995fc,Hart:1996wk,Fukushima:1997yn,Fukushima:1999gf,Brower:1997js}
This correlation suggests that instantons are important 
for the multi-production of monopole loops.\cite{Fukushima:1997yn,Fukushima:1999gf}
For this reason, we investigate further the background-field 
effect on the appearance of monopoles around instantons.

The monopole current is extracted as follows. We use
the MA gauge,\cite{Kronfeld:1987vd} 
which is defined in the $SU(2)$ case by minimizing 
\begin{equation}
R_{ch}  =  2 \sum_{s,\mu} \Bigm[1-\frac{1}{2} \{ 
(U_{\mu}^{1}(s))^{2}+(U_{\mu}^{2}(s))^{2}\} \Bigm],
\label{EMAL}
\end{equation}
with $U_{\mu}\!=\!U_{\mu}^{0}\!+\!i\tau^{i}U_{\mu}^{i}$.
In the MA gauge, the link variable is decomposed as
$U_{\mu}(s) = M_{\mu}(s) u_{\mu}(s)$, with 
\begin{eqnarray}
\!\!M_{\mu}(s) \equiv \!
\left(\!\!\!\!
        \begin{array}{cc}
\sqrt{1-\!\mid\!c_{\mu}(s)\!\mid^{2}} &
            \!\!\!- c^{*}_{\mu}(s) \\
             c_{\mu}(s)          & 
\!\!\!\sqrt{1-\!\mid\!c_{\mu}(s)\!\mid^{2}}
        \end{array}
  \!\!\right),~
u_{\mu}(s) \equiv\!
\left(\!\!
        \begin{array}{cc}
   e^{i\theta_{\mu}(s)} &
    \!\!        0           \\
            0             
  & \!\! e^{-i\theta_{\mu}(s)}      
        \end{array}
  \!\!\!\right),~~~
\end{eqnarray}
where the Abelian angle variable $\theta_{\mu}$ and 
the non-Abelian variable $c_{\mu}$ are defined in 
terms of $U_{\mu}$ as $\tan\theta_{\mu} = U_{\mu}^{3}
/U_{\mu}^{0}$ and $c_{\mu}e^{i\theta_{\mu}} 
=-U_{\mu}^{2} + i U_{\mu}^{1}$. It is obvious from 
the expression of Eq.~(\ref{EMAL}) that the off-diagonal 
parts $U_{\mu}^{1}$ and $U_{\mu}^{2}$ of the gluon fields 
are minimized in the MA gauge. Therefore, full 
$SU(2)$ link variables are approximated as 
$U(1)$ link variables, $U_{\mu} \simeq u_{\mu}$, 
in the MA gauge. 

Monopole currents are defined by $u_{\mu}(s)$, following 
DeGrand and Toussaint.\cite{DeGrand:1980eq} 
Using the forward derivative $\partial_{\mu}f(s)\equiv f(s+\hat{\mu})
-f(s)$ with unit vector $\hat{\mu}$, the 2-form 
of the lattice formulation, $\theta_{\mu\nu}(s) 
\equiv \partial_{\mu}\theta_{\nu}(s)-\partial_{\nu}
\theta_{\mu}(s)$, is decomposed as
\begin{equation}
\theta_{\mu\nu}(s) = \bar{\theta}_{\mu\nu}(s) 
+ 2\pi n_{\mu\nu}(s),
\end{equation}
with $\bar{\theta}_{\mu\nu}(s) \equiv {\rm mod}_{2\pi}
\theta_{\mu\nu} \in(-\pi, \pi]$ 
and $n_{\mu\nu}(s) \in \mib{Z}$. Here, 
$\bar{\theta}_{\mu\nu}(s)$ and $2\pi n_{\mu\nu}(s)$ 
correspond to the regular field strength and the singular 
Dirac string part, respectively. Since the Abelian Bianchi 
identity is broken, the monopole current $k_{\mu}(^*\!s)$ 
appears on the dual link $(^*\!s,\mu)$ as
\begin{equation}
k_{\mu}(^*\!s) \equiv (1/4\pi)
\varepsilon_{\mu\nu\alpha\beta}
\partial_{\nu}\bar{\theta}_{\alpha\beta}(s+\hat{\mu}) 
=  - \partial_{\nu}\tilde{n}_{\mu\nu}(^*\!s),
\end{equation}
where $\tilde{n}_{\mu\nu}(s) \equiv \frac{1}{2}
\varepsilon_{\mu\nu\alpha\beta}n_{\alpha\beta}
(s+\hat{\mu})$. The current-conservation 
law $\partial^{\prime}_{\mu}k_{\mu}(^*\!s) = 0$
leads to a closed monopole loop in $\mib{R}^{4}$. 
Here, $\partial^{\prime}_{\mu}$ 
denotes a backward derivative. 

Here, the self-dual solution automatically satisfies 
the MA gauge local condition $(\partial_{\mu} \mp 
A^{3}_{\mu})A^{\pm}_{\mu}=0$, which gives a stationary 
point of the functional $R_{ch}$. For the case of a 
single instanton on the lattice, the monopole loop is 
localized around the center of the instanton. However, 
such a monopole-loop plane cannot be determined, 
because there is no definite direction in either
the single instanton configuration or in the MA gauge 
condition, due to 4-dim rotational invariance. 
In this respect, the external field plays an important
role in the determination of the monopole trajectory
formed by the instanton. Due to the homogeneous external 
field, the $O(4)$ rotational symmetry is reduced to 
$O(2)\!\times\!O(2)$, which specifies the 2-dim plane 
of the external field, $F_{12}\ne0$. In fact, there appears 
an asymmetry between the $xy$ plane and the $zt$ plane 
for the $F_{12}\!\ne\!0$ case.

\begin{figure}[b]
\parbox{\halftext}{
\epsfxsize = 5.5cm
\centerline{\epsfbox{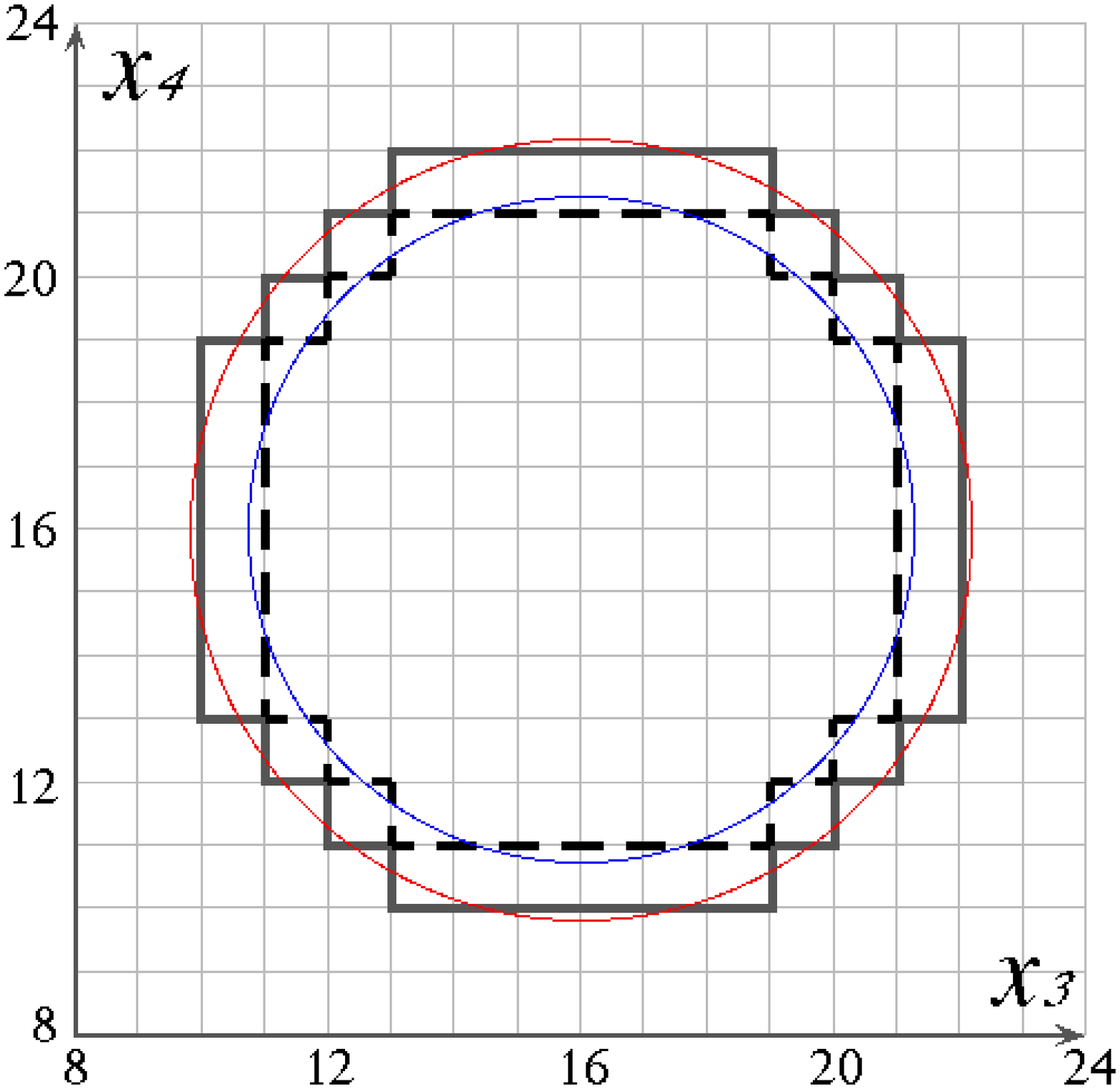}}
\caption{The monopole loop around an instanton 
located at the lattice center in an external field 
after the action minimization procedure has been 
applied. Here, we use the MA gauge. The dashed and 
solid curves denote 
monopole loops for $F_{12}/4\pi\!=\!1$ and $\!\sqrt{3}
~{\rm fm^{-2}}$, corresponding to $\langle\! 
G^{a}_{\!\mu\nu} G^{a}_{\!\mu\nu}\!\rangle /32\pi^{2}\!\!=\!1$ 
and $3~{\rm fm^{\!-4}}$, respectively.}
\label{FMC}
}
\hfill
\parbox{\halftext}{
\epsfxsize = 6.6cm
\centerline{\epsfbox{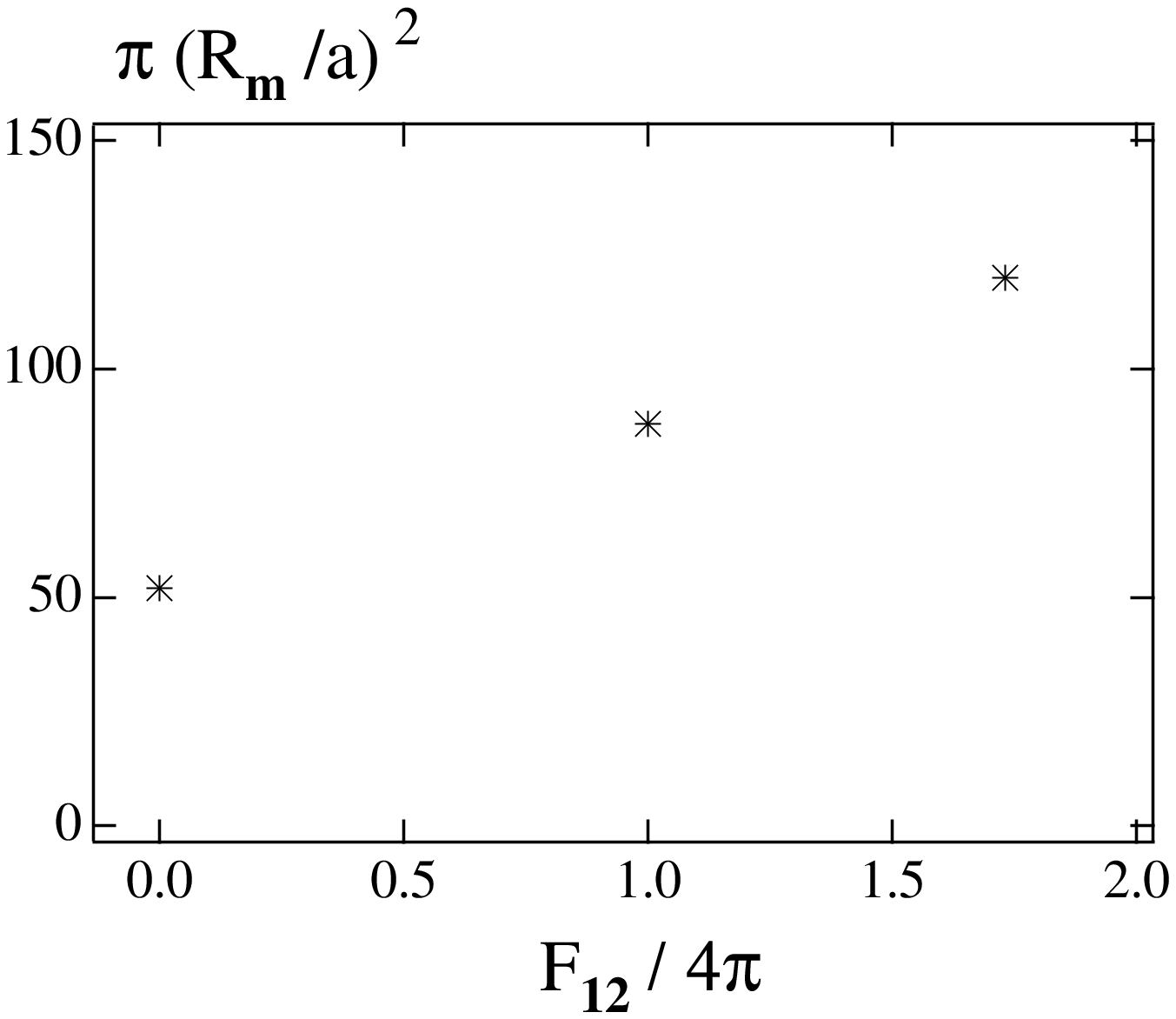}}
\vspace*{-0.3cm}
\caption{The dependence of the minimal area enclosed by
the monopole loop on the strength of the 
external field $F_{12}/4\pi$. Here, $R_{m}$ is the radius
of the monopole loop.}
\label{FMCA}
}
\end{figure}

We now apply the MA gauge fixing to the superposed gauge field 
of the external field and the stabilized instanton that was
obtained in the previous section and extract monopole 
currents. Then, there appears a local correlation between 
the instanton and monopole. In an external field with $F_{12} 
> 0$, a monopole loop appears around the instanton 
configuration in the $zt$ plane, including its center,
as shown in Fig.~\ref{FMC}. In other words, pair creation of 
a monopole and an anti-monopole occurs at a time $t$ along 
the external field $H_{3}\ne 0$, and after some time interval
pair annihilation occurs, which seems to partially screen 
the external magnetic field. Figure~\ref{FMCA} displays the 
dependence of the closed area of the monopole loop on the 
strength of the external field. 
The external field plays a significant role in the 
promotion of large monopole-loops.

\section{Summary and conclusion}
\label{sec:sec7}
In this paper, we have investigated a deformation 
mechanism of the instanton configuration in the 
presence of a homogeneous external field. 
An isolated single instanton has global color 
orientations $O^{ab}$ as collective coordinates
and gauge degrees of freedom $\Omega(x)$. However,
external fields lead to the dependence of the
total Yang-Mills action on these variables and the
breaking of the $O(4)$ symmetry of the instanton 
solution. In this situation, the 't Hooft ansatz 
is not a solution of the Yang-Mills field equation 
and is unstable in the variational 
space of the instanton configuration, whose degrees
of freedom are the global color orientation $O^{ab}$, 
the anisotropic scale $\lambda$, and the local 
gauge-like function $\Omega(x)$.
In this way, we have found the stable instanton 
solution that minimizes the total 
Yang-Mills action starting from a total gauge 
field that is a superposition of the 't Hooft 
ansatz in the singular gauge with an external field.

We have investigated the case of a homogeneous external 
field $F_{12}> 0$ for simplicity and determined $O^{ab}$, 
$\theta$ and $\Omega(x)$ so as to minimize the total 
action. Due to the external field, the $O(4)$ symmetry 
is reduced to two 2-dimensional rotational symmetries 
$O(2)\times 
O(2)$ in the $xy$ and $zt$ planes in the instanton solution. 
In the space $\mib{R}^{3}$ at fixed $t$, we found a 
quadrupole deformation of this instanton solution. In 
a magnetic field $\vec{H}$, a prolate deformation 
occurs along the direction of $\vec{H}$. 
The magnitude of the deformation is found to increase 
with the instanton size and the strength of background 
fields. Since the QCD vacuum is composed of many 
instantons and anti-instantons, instanton configurations 
seem to lose the $O(4)$ symmetry and become deformed 
by such background fields.

We have considered further the effect of external fields
on the local correlation between instantons and monopoles
in the MA gauge. 
Both for the single instanton and MA gauge conditions,
there is no definite direction in $\mib{R}^{4}$, and therefore 
in the absence of a background field, the direction of a 
monopole loop is ``fragile''. A homogeneous
external field breaks the 4-dimensional rotational 
invariance on the instanton configuration, and as a result,
a monopole trajectory around the instanton is formed with 
a direction corresponding to the external field. A monopole
and anti-monopole pair is created at a time $t$ along the
$z$ direction of the external field $H_{3}\ne 0$
and after some time interval 
pair annihilation occurs in a $zt$ plane. In fact, such a 
monopole loop seems to partially screen the external field. 
The closed minimal area of the monopole-loop generated by a 
fixed size instanton is found to increase with the strength 
of the external field. 

\section*{Acknowledgements}
\label{sec:sec8}
We would like to thank Dr.~H.~Toki for his useful comments and
discussions. We have carried out all numerical simulations 
reported in this paper on a NEC SX5 at RCNP.

\appendix
\section{}

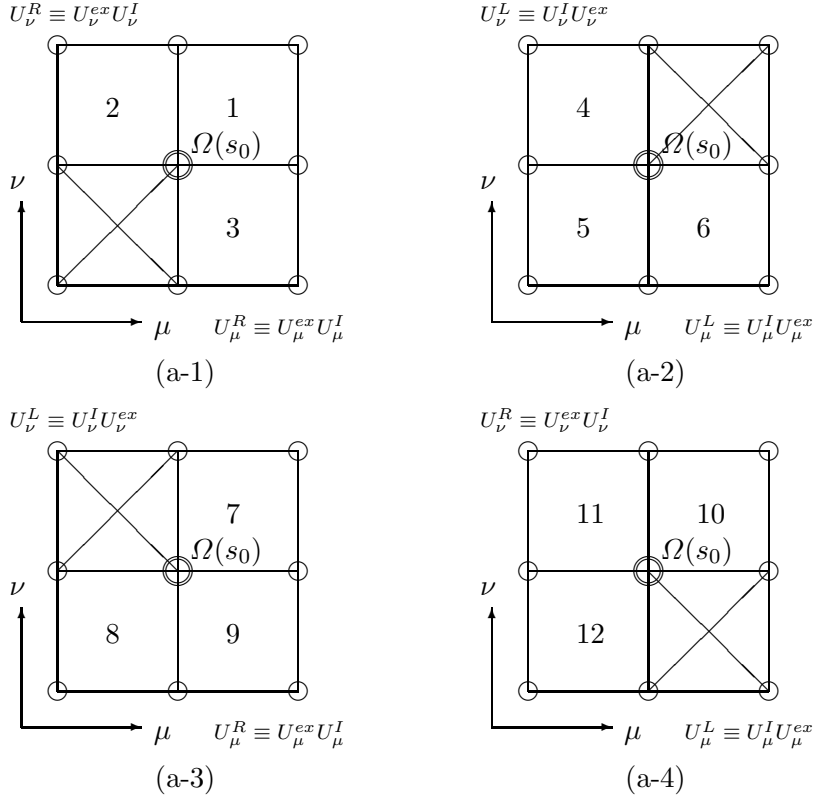
\begin{figure}[b]
\begin{center}
\begin{minipage}[hbt]{5.0cm}
\setlength{\unitlength}{1.6mm}
\begin{picture} (30,30) (0,0)
\multiput(5,5)(0,10){3}{\line(1,0){20}}
\multiput(5,5)(10,0){3}{\line(0,1){20}}
\multiput(5,5)(10,0){3}{\circle{1.5}}
\multiput(5,15)(20,0){2}{\circle{1.5}}
\multiput(5,25)(10,0){3}{\circle{1.5}}
\put(15,15){\circle{2.5}}
\put(15,15){\circle{2}}
\put(16,16){$\Omega(s_{0})$}
\put(2,2) {\vector(0,1){10}}
\put(2,2) {\vector(1,0){10}} 
\put(1,13) {$\nu$}
\put(13,1) {$\mu$}
\put(19,19) {$1$}
\put(9,19) {$2$}
\put(19,9) {$3$}
\put(5,5){\line(1,1){10}}
\put(5,15){\line(1,-1){10}}
\put(1,27) {\scriptsize $U^{R}_{\nu}\equiv U^{ex}_{\nu}U^{I}_{\nu}$}
\put(18,1) {\scriptsize $U^{R}_{\mu}\equiv U^{ex}_{\mu}U^{I}_{\mu}$}
\end{picture}

\center{(a-1)}
\end{minipage}
\hspace{1cm}
\begin{minipage}[hbt]{5.0cm}
\setlength{\unitlength}{1.6mm}
\begin{picture} (30,30) (0,0)
\multiput(5,5)(0,10){3}{\line(1,0){20}}
\multiput(5,5)(10,0){3}{\line(0,1){20}}
\multiput(5,5)(10,0){3}{\circle{1.5}}
\multiput(5,15)(20,0){2}{\circle{1.5}}
\multiput(5,25)(10,0){3}{\circle{1.5}}
\put(15,15){\circle{2.5}}
\put(15,15){\circle{2}}
\put(16,16){$\Omega(s_{0})$}
\put(2,2) {\vector(0,1){10}}
\put(2,2) {\vector(1,0){10}} 
\put(1,13) {$\nu$}
\put(13,1) {$\mu$}
\put(9,19) {$4$}
\put(9, 9) {$5$}
\put(19,9) {$6$}
\put(15,15){\line(1,1){10}}
\put(15,25){\line(1,-1){10}}
\put(1,27) {\scriptsize $U^{L}_{\nu}\equiv U^{I}_{\nu}U^{ex}_{\nu}$}
\put(18,1) {\scriptsize $U^{L}_{\mu}\equiv U^{I}_{\mu}U^{ex}_{\mu}$}
\end{picture}

\center{(a-2)}
\end{minipage}

\begin{minipage}[hbt]{5.0cm}
\setlength{\unitlength}{1.6mm}
\begin{picture} (30,30) (0,0)
\multiput(5,5)(0,10){3}{\line(1,0){20}}
\multiput(5,5)(10,0){3}{\line(0,1){20}}
\multiput(5,5)(10,0){3}{\circle{1.5}}
\multiput(5,15)(20,0){2}{\circle{1.5}}
\multiput(5,25)(10,0){3}{\circle{1.5}}
\put(15,15){\circle{2.5}}
\put(15,15){\circle{2}}
\put(16,16){$\Omega(s_{0})$}
\put(2,2) {\vector(0,1){10}}
\put(2,2) {\vector(1,0){10}} 
\put(1,13) {$\nu$}
\put(13,1) {$\mu$}
\put(19,19) {$7$}
\put(9,9) {$8$}
\put(19,9) {$9$}
\put(5,15){\line(1,1){10}}
\put(5,25){\line(1,-1){10}}
\put(1,27) {\scriptsize $U^{L}_{\nu}\equiv U^{I}_{\nu}U^{ex}_{\nu}$}
\put(18,1) {\scriptsize $U^{R}_{\mu}\equiv U^{ex}_{\mu}U^{I}_{\mu}$}
\end{picture}

\center{(a-3)}
\end{minipage}
\hspace{1cm}
\begin{minipage}[hbt]{5.0cm}
\setlength{\unitlength}{1.6mm}
\begin{picture} (30,30) (0,0)
\multiput(5,5)(0,10){3}{\line(1,0){20}}
\multiput(5,5)(10,0){3}{\line(0,1){20}}
\multiput(5,5)(10,0){3}{\circle{1.5}}
\multiput(5,15)(20,0){2}{\circle{1.5}}
\multiput(5,25)(10,0){3}{\circle{1.5}}
\put(15,15){\circle{2.5}}
\put(15,15){\circle{2}}
\put(16,16){$\Omega(s_{0})$}
\put(2,2) {\vector(0,1){10}}
\put(2,2) {\vector(1,0){10}} 
\put(1,13) {$\nu$}
\put(13,1) {$\mu$}
\put(19,19) {$10$}
\put(9,19) {$11$}
\put(9,9) {$12$}
\put(15,5){\line(1,1){10}}
\put(15,15){\line(1,-1){10}}
\put(1,27) {\scriptsize $U^{R}_{\nu}\equiv U^{ex}_{\nu}U^{I}_{\nu}$}
\put(18,1) {\scriptsize $U^{L}_{\mu}\equiv U^{I}_{\mu}U^{ex}_{\mu}$}
\end{picture}

\center{(a-4)}
\end{minipage}
\caption{Wilson plaquette around a site $s_{0}$. Here, we use
the gauge transformation $\Omega(s_{0})$ of $s_{0}$. To maintain
the ordering symmetry for the product of $U^{I}_{\mu}$  and 
$U^{ex}_{\mu}$, we have to consider both of the ordering choices 
$U^{R}=U^{ex}U^{I}$ and $U^{L}=U^{I}U^{ex}$ for each direction
$\mu$ and $\nu$.
Therefore, we consider four types of clover plaquettes, 
(1),(2),(3) and (4), which correspond to (RR), (LL), (RL) 
and (LR), respectively. The cross-checked plaquette in each 
clover is independent of the local gauge transformation of 
$\Omega(s_0)$, and thus does not contribute to 
$\delta S_{YM}(s_0;\Omega(s_0))$.}
\label{fig:PLQSYM} 
\end{center}
\end{figure}

We must maintain the geometrical symmetry and the 
ordering symmetry for the product of $U_{\mu}^{I}$ 
and $U_{\mu}^{ex}$ in the minimization procedure.
To maintain the ordering symmetry for the product of $U^{I}$  
and $U^{ex}$, we have to consider both of the ordering choices 
$U^{R}=U^{ex}U^{I}$ and $U^{L}=U^{I}U^{ex}$, for each 
direction $\mu$ and $\nu$. As shown in Fig.~\ref{fig:PLQSYM}, 
we therefore consider four types of plaquettes (1),(2),(3) 
and (4), which correspond to (RR), (LL), (RL) and (LR), 
respectively. Here, we use the clover-type plaquettes 
because of the geometrical symmetry. The cross-checked plaquette 
in each clover is independent of the local gauge 
transformation of $\Omega(s_0)$ and does not contribute 
to $\delta S_{YM}(s_0;\Omega(s_0))$.
We list 12 components of $\hat{L}^{\alpha}_{\mu\nu}$ 
and $\bar{L}^{\alpha}_{\mu\nu}$ in Eq.~(\ref{DIFFAC}) 
for the construction of $\delta S_{YM}$ as follows.

The RR part is written 
\begin{eqnarray}
\hat{L}^{1}_{\mu\nu}(s_{0})&=&
U^{ex\dagger}_{\nu}(s_{0})U^{ex}_{\mu}(s_{0}), \nonumber\\
\bar{L}^{1}_{\mu\nu}(s_{0})&=&
U^{I}_{\mu}(s_{0})U^{R}_{\nu}(s_{0}+\hat{\mu})
U^{R\dagger}_{\mu}(s_{0}+\hat{\nu})U^{I\dagger}_{\nu}(s_{0}),\nonumber\\
\hat{L}^{2}_{\mu\nu}(s_{0})&=&
U^{ex}_{\nu}(s_{0}),\nonumber\\
\bar{L}^{2}_{\mu\nu}(s_{0})&=&
U^{I}_{\nu}(s_{0})
U^{R\dagger}_{\mu}(s_{0}-\hat{\mu}+\hat{\nu})
U^{R\dagger}_{\nu}(s_{0}-\hat{\mu})
U^{R}_{\mu}(s_{0}-\hat{\mu}),\nonumber\\
\hat{L}^{3}_{\mu\nu}(s_{0})&=&
U^{ex\dagger}_{\mu}(s_{0}),\nonumber\\
\bar{L}^{3}_{\mu\nu}(s_{0})&=&
U^{R\dagger}_{\nu}(s_{0}-\hat{\nu})
U^{R}_{\mu}(s_{0}-\hat{\nu})
U^{R}_{\nu}(s_{0}+\hat{\mu}-\hat{\nu})
U^{I\dagger}_{\mu}(s_{0}).
\end{eqnarray}

The LL part is written as 
\begin{eqnarray}
\hat{L}^{4}_{\mu\nu}(s_{0})&=&
U^{ex}_{\mu}(s_{0}-\hat{\mu}), \nonumber\\
\bar{L}^{4}_{\mu\nu}(s_{0})&=&
U^{L}_{\nu}(s_{0})
U^{L\dagger}_{\mu}(s_{0}-\hat{\mu}+\hat{\nu})
U^{L\dagger}_{\nu}(s_{0}-\hat{\mu})
U^{I}_{\mu}(s_{0}-\hat{\mu}),\nonumber\\
\hat{L}^{5}_{\mu\nu}(s_{0})&=&
U^{ex}_{\nu}(s_{0}-\hat{\nu})
U^{ex\dagger}_{\mu}(s_{0}-\hat{\mu}),\nonumber\\
\bar{L}^{5}_{\mu\nu}(s_{0})&=&
U^{I\dagger}_{\mu}(s_{0}-\hat{\mu})
U^{L\dagger}_{\nu}(s_{0}-\hat{\mu}-\hat{\nu})
U^{L}_{\mu}(s_{0}-\hat{\mu}-\hat{\nu})
U^{I}_{\nu}(s_{0}-\hat{\nu}),~~\nonumber\\
\hat{L}^{6}_{\mu\nu}(s_{0})&=&
U^{ex\dagger}_{\nu}(s_{0}-\hat{\nu}),~~\nonumber\\
\bar{L}^{6}_{\mu\nu}(s_{0})&=&
U^{I\dagger}_{\nu}(s_{0}-\hat{\nu})
U^{L}_{\mu}(s_{0}-\hat{\nu})
U^{L}_{\nu}(s_{0}+\hat{\mu}-\hat{\nu})
U^{L\dagger}_{\mu}(s_{0}).
\end{eqnarray}

The RL part is written as 
\begin{eqnarray}
\hat{L}^{7}_{\mu\nu}(s_{0})&=&
U^{ex}_{\mu}(s_{0}), \nonumber\\
\bar{L}^{7}_{\mu\nu}(s_{0})&=&
U^{I}_{\mu}(s_{0})
U^{L}_{\nu}(s_{0}+\hat{\mu})
U^{R\dagger}_{\mu}(s_{0}+\hat{\nu})
U^{L\dagger}_{\nu}(s_{0}),\nonumber\\
\hat{L}^{8}_{\mu\nu}(s_{0})&=&
U^{ex}_{\nu}(s_{0}-\hat{\nu}),\nonumber\\
\bar{L}^{8}_{\mu\nu}(s_{0})&=&
U^{R\dagger}_{\mu}(s_{0}-\hat{\mu})
U^{L\dagger}_{\nu}(s_{0}-\hat{\mu}-\hat{\nu})
U^{R}_{\mu}(s_{0}-\hat{\mu}-\hat{\nu})
U^{I}_{\nu}(s_{0}-\hat{\nu}),\nonumber\\
\hat{L}^{9}_{\mu\nu}(s_{0})&=&
U^{ex\dagger}_{\mu}(s_{0})
U^{ex\dagger}_{\nu}(s_{0}-\hat{\nu}),\nonumber\\
\bar{L}^{9}_{\mu\nu}(s_{0})&=&
U^{I\dagger}_{\nu}(s_{0}-\hat{\nu})
U^{R}_{\mu}(s_{0}-\hat{\nu})
U^{L}_{\nu}(s_{0}+\hat{\mu}-\hat{\nu})
U^{I\dagger}_{\mu}(s_{0}).
\end{eqnarray}

The LR part is written as 
\begin{eqnarray}
\hat{L}^{10}_{\mu\nu}(s_{0})&=&
U^{ex\dagger}_{\nu}(s_{0}),\nonumber\\
\bar{L}^{10}_{\mu\nu}(s_{0})&=&
U^{L}_{\mu}(s_{0})
U^{R}_{\nu}(s_{0}+\hat{\mu})
U^{L\dagger}_{\mu}(s_{0}+\hat{\nu})
U^{I\dagger}_{\nu}(s_{0}),\nonumber\\
\hat{L}^{11}_{\mu\nu}(s_{0})&=&
U^{ex}_{\mu}(s_{0}-\hat{\mu})
U^{ex}_{\nu}(s_{0}),\nonumber\\
\bar{L}^{11}_{\mu\nu}(s_{0})&=&
U^{I}_{\nu}(s_{0})
U^{L\dagger}_{\mu}(s_{0}-\hat{\mu}+\hat{\nu})
U^{R\dagger}_{\nu}(s_{0}-\hat{\mu})
U^{I}_{\mu}(s_{0}-\hat{\mu}),\nonumber\\
\hat{L}^{12}_{\mu\nu}(s_{0})&=&
U^{ex\dagger}_{\mu}(s_{0}-\hat{\mu}),\nonumber\\
\bar{L}^{12}_{\mu\nu}(s_{0})&=&
U^{I\dagger}_{\mu}(s_{0}-\hat{\mu})
U^{R\dagger}_{\nu}(s_{0}-\hat{\mu}-\hat{\nu})
U^{L}_{\mu}(s_{0}-\hat{\mu}-\hat{\nu})
U^{R}_{\nu}(s_{0}-\hat{\nu}).
\end{eqnarray}


\begin{thebibliography}{99}
\expandafter\ifx\csname bibnamefont\endcsname\relax
  \def\bibnamefont#1{#1}\fi
\expandafter\ifx\csname bibfnamefont\endcsname\relax
  \def\bibfnamefont#1{#1}\fi
\expandafter\ifx\csname url\endcsname\relax
  \def\url#1{\texttt{#1}}\fi
\expandafter\ifx\csname urlprefix\endcsname\relax\def\urlprefix{URL }\fi
\expandafter\ifx\csname bibinfo\endcsname\relax \def\bibinfo#1#2{#2}\fi
\expandafter\ifx\csname eprint\endcsname\relax \def\eprint#1{#1}\fi

\bibitem{Belavin:1975fg}
A.~A.~Belavin, A.~M.~Polyakov, A.~S.~Schwartz and Y.~S.~Tyupkin, Phys. Lett. B
{\bf 59} (1975), 85.

\bibitem{'tHooft:1976up}
G.~'t~Hooft, Phys. Rev. Lett. {\bf 37} (1976), 8.

\bibitem{Veneziano:1979ec}
G.~Veneziano, Nucl. Phys. B {\bf 159} (1979), 213.

\bibitem{Witten:1979vv}
E.~Witten, Nucl. Phys. B {\bf 156} (1979), 269. 

\bibitem{Shuryak:1990cx}
E.~V.~Shuryak and J.~J.~M.~Verbaarschot, Nucl. Phys. B {\bf 341}
  (1990), 1.

\bibitem{Schafer:1998wv}
T.~Schafer and E.~V.~Shuryak, Rev. Mod. Phys. {\bf 70} (1998), 323.

\bibitem{Diakonov:1995ea}
D.~Diakonov, hep-ph/9602375.

\bibitem{Teper:1985rb}
M.~Teper, Phys. Lett. B {\bf 162} (1985), 357.

\bibitem{Ilgenfritz:1986dz} E.~M.~Ilgenfritz, M.~L.~Laursen, G.~Schierholz,
M.~Muller-Preussker and H.~Schiller, Nucl. Phys. B {\bf 268} (1986), 693.

\bibitem{Polikarpov:1988yr}
M.~I.~Polikarpov and A.~I.~Veselov, Nucl. Phys. B {\bf 297}
 (1988), 34.

\bibitem{Campostrini:1990dh}
M.~Campostrini, A.~D.~Giacomo, H. Panagopoulos and E. Vicari,
Nucl. Phys. B {\bf 329} (1990), 683.

\bibitem{Teper:1994un}
M.~Teper, Nucl. Phys. B {\bf 411} (1994), 855.

\bibitem{Michael:1995br}
C.~Michael and P.~S.~Spencer, Phys. Rev. D {\bf 52}
(1995), 4691.

\bibitem{deForcrand:1997sq}
P.~de~Forcrand, M.~G.~Perez and I.-O.~Stamatescu, Nucl. Phys. B {\bf 499}
  (1997), 409.

\bibitem{DeGrand:1997ss}
T.~DeGrand, A.~Hasenfratz and T.~G.~Kovacs, Nucl. Phys. B {\bf 520}
 (1998), 301.

\bibitem{Savvidy:1977as}
G.~K.~Savvidy, Phys. Lett. B {\bf 71} (1977), 133.

\bibitem{Nielsen:1978rm}
N.~K.~Nielsen and P.~Olesen, Nucl. Phys. B {\bf 144} (1978), 376. 

\bibitem{Ambjorn:1980ms}
J. Ambjorn and P.~Olesen, Nucl. Phys. B {\bf 170}
 (1980), 265.

\bibitem{Ambjorn:1980xi}
J.~Ambjorn and P.~Olesen, Nucl. Phys. B {\bf 170} (1980), 60.

\bibitem{Negele:1998ev}
J.~W.~Negele, Nucl. Phys. Proc. Suppl. {\bf 73} (1999), 92.

\bibitem{Suganuma:1996qr}
H.~Suganuma, A.~Tanaka, S.~Sasaki and O.~Miyamura,
 Nucl. Phys. Proc. Suppl. {\bf 47} (1996), 302.

\bibitem{Miyamura:1995fc}
O.~Miyamura and S.~Origuchi, {\it Confinement '95}
  (World Scientific, 1995), p.~235.

\bibitem{Hart:1996wk}
A.~Hart and M.~Teper, Phys. Lett. B {\bf 371}
 (1996), 261.

\bibitem{Fukushima:1997yn}
M.~Fukushima, S.~Sasaki, H.~Suganuma, A.~Tanaka, H.~Toki 
and D.~Diakonov, Phys. Lett. B {\bf 399} (1997), 141.

\bibitem{Fukushima:1999gf}
M.~Fukushima, H.~Suganuma and H.~Toki, Phys. Rev. D {\bf 60}
(1999), 094504.

\bibitem{Brower:1997js}
R.~C.~Brower, K.~N.~Orginos and C.-I.~Tan, Phys. Rev. D {\bf 55}
(1997), 6313.

\bibitem{Rajaraman:1982bk}
R. Rajaraman, {\it Solitons~and~Instanton} (North-Holland, 1982), p.~1.

\bibitem{Shuryak:1988bk}
E.~Shuryak, {\it The QCD Vacuum, Hadron and The Superdense Matter}
  (World Scientific, 1988), p.~1.

\bibitem{Munster:2000uu}
G.~Munster and C.~Kamp, Eur. Phys. J. C {\bf 17} (2000), 447.



\bibitem{'tHooft:1981ht}
G.~'t~Hooft, Nucl. Phys. B {\bf 190} (1981), 455. 

\bibitem{Nambu:1974pb}
Y.~Nambu and M.~Y.~Han, Phys. Rev. D {\bf 10} (1974), 674.

\bibitem{Nambu:1974zg}
Y.~Nambu, Phys. Rev. D {\bf 10} (1974), 4262.

\bibitem{'tHooft:1975pu}
G.~'t~Hooft, in {\it High Energy Physics, Proceedings of European Physical Society International Conference, Palermo, Italy, Jun 23-28, 1975}.

\bibitem{Mandelstam:1974pi}
S.~Mandelstam, Phys. Rep. {\bf 23} (1976), 245. 

\bibitem{Kronfeld:1987vd}
A.~S.~Kronfeld, G.~Schierholz'and U.~J.~Wiese,
Nucl. Phys. B {\bf 293} (1987), 461. 

\bibitem{Kronfeld:1987ri}
A.~S.~Kronfeld, M.~L.~Laursen, G.~Schierholz and U.~J.~Wiese, Phys. Lett. B
{\bf 198} (1987), 516. 

\bibitem{Brandstater:1991sn}
F.~Brandstater, U.~J.~Wiese and G.~Schierholz, Phys. Lett. B {\bf 272}
(1991), 319. 

\bibitem{Hioki:1991ai}
S.~Hioki, S.~Kitahara, S.~Kiura, Y.~Matsubara, O.~Miyamura, S.~Ohno and
T.~Suzuki, Phys. Lett. B {\bf 272}
  (1991), 326.

\bibitem{Kitahara:1995vt}
S.~Kitahara, Y.~Matsubara and T.~Suzuki, Prog. Theor. Phys. {\bf 93}
(1995), 1.

\bibitem{Miyamura:1995xn}
O.~Miyamura, Phys. Lett. B {\bf 353} (1995), 91. 

\bibitem{Miyamura:1994xk}
O.~Miyamura, Nucl. Phys. Proc. Suppl. {\bf 42}
(1995), 538. 

\bibitem{Woloshyn:1995rv}
R.~M.~Woloshyn, Phys. Rev. D {\bf 51} (1995), 6411.

\bibitem{Suganuma:1995ps}
H.~Suganuma, S.~Sasaki and H.~Toki,
Nucl. Phys. B {\bf 435} (1995), 207.

\bibitem{Sasaki:1995sa}
S.~Sasaki, H.~Suganuma and H.~Toki, Prog. Theor. Phys. {\bf 94}
 (1995), 373.

\bibitem{Sasaki:1996ss}
S.~Sasaki, H.~Suganuma and H.~Toki, Phys. Lett. B {\bf 387} (1996), 145.

\bibitem{DeGrand:1980eq}
T.~A.~DeGrand and D.~Toussaint, Phys. Rev. D {\bf 22}
 (1980), 2478.

\end{thebibliography}
\end{document}